\documentclass[aps,pra,reprint,superscriptaddress,longbibliography]{revtex4-1}
\usepackage[utf8]{inputenc}
\usepackage{color}
\usepackage{graphicx}
\usepackage{epstopdf}
\usepackage{siunitx}
\usepackage{amsmath}
    \DeclareSIUnit{\wtpercent}{wt\%}
\usepackage{upgreek}

\usepackage[capitalise]{cleveref}
\newcommand{\MassFracW}{w}
\newcommand{\VapConcW}{c}
\newcommand{\VapConcWVLE}{c_\mathrm{VLE}}
\newcommand{\VapConcWInfty}{c_\infty}
\newcommand{\MassTransW}{j}
\newcommand{\PChamber}{P_\mathrm{ch}}
\newcommand{\dotPChamber}{\dot{P}_\mathrm{ch}}
\newcommand{\Compliance}{C_\mathrm{ch}}
\newcommand{\VoltToVolumeFactor}{\alpha}

\newcommand{\dotV}{\dot{V}}

\usepackage[normalem]{ulem}

\begin{document}

\title{Selective evaporation at the nozzle exit in piezoacoustic inkjet printing}

\author{Maaike Rump}
\affiliation{Physics of Fluids group, Max Planck Center Twente for Complex Fluid Dynamics and J. M. Burgers Center for Fluid Dynamics, University of Twente, 7500AE Enschede, The Netherlands}
\author{Uddalok Sen}
\affiliation{Physics of Fluids group, Max Planck Center Twente for Complex Fluid Dynamics and J. M. Burgers Center for Fluid Dynamics, University of Twente, 7500AE Enschede, The Netherlands}
\affiliation{Physical Chemistry and Soft Matter, Wageningen University and Research, 6708 WE Wageningen, The Netherlands}
\author{Roger Jeurissen}
\affiliation{Physics of Fluids group, Max Planck Center Twente for Complex Fluid Dynamics and J. M. Burgers Center for Fluid Dynamics, University of Twente, 7500AE Enschede, The Netherlands}
\author{Hans Reinten}
\affiliation{Physics of Fluids group, Max Planck Center Twente for Complex Fluid Dynamics and J. M. Burgers Center for Fluid Dynamics, University of Twente, 7500AE Enschede, The Netherlands}
\author{Michel Versluis}
\affiliation{Physics of Fluids group, Max Planck Center Twente for Complex Fluid Dynamics and J. M. Burgers Center for Fluid Dynamics, University of Twente, 7500AE Enschede, The Netherlands}
\author{Detlef Lohse}
\affiliation{Physics of Fluids group, Max Planck Center Twente for Complex Fluid Dynamics and J. M. Burgers Center for Fluid Dynamics, University of Twente, 7500AE Enschede, The Netherlands}
\affiliation{Max Planck Institute for Dynamics and Self-Organization, Am Fa{\ss}berg 17, 37077 Göttingen, Germany}
\author{Christian Diddens}
\affiliation{Physics of Fluids group, Max Planck Center Twente for Complex Fluid Dynamics and J. M. Burgers Center for Fluid Dynamics, University of Twente, 7500AE Enschede, The Netherlands}
\author{Tim Segers}
\affiliation{BIOS / Lab on a Chip Group, Max Planck Center Twente for Complex Fluid Dynamics, MESA+ Institute for Nanotechnology, University of Twente, 7500AE Enschede, The Netherlands}

\begin{abstract}
In practical applications of inkjet printing the nozzles in a printhead have intermittent idle periods, during which ink can evaporate from the nozzle exit. Inks are usually multicomponent where each component has its own characteristic evaporation rate resulting in concentration gradients within the ink. These gradients may directly and indirectly (via Marangoni flows) alter the jetting process and thereby its reproducibility and the resulting print quality. In the present work, we study selective evaporation from an inkjet nozzle for water-glycerol mixtures. Through experiments, analytical modeling, and numerical simulations, we investigate changes in mixture composition with drying time. By monitoring the acoustics within the printhead, and subsequently modeling the system as a mass-spring-damper system, the composition of the mixture can be obtained as a function of drying time. The results from the analytical model are validated using numerical simulations of the full fluid mechanical equations governing the printhead flows and pressure fields. Furthermore, the numerical simulations reveal that the time independent concentration gradient we observe in the experiments is due to the steady state of water flux through the printhead. Finally, we measure the number of drop formation events required in this system before the mixture concentration within the nozzle attains the initial (pre-drying) value, and find a stronger than exponential trend in the number of drop formations required. These results shed light on the complex physiochemical hydrodynamics associated with the drying of ink at a printhead nozzle, and help in increasing the stability and reproducibility of inkjet printing.
\end{abstract}

\maketitle

\section{Introduction}

Modern inkjet printing {\color{red} \sout{for complex graphics}} employs drop-on-demand (DOD) printheads as these have major advantages over continuous inkjet (CIJ) printers including the fact that there is no need for complicated hardware and complex electronic circuitry for jet break-up synchronization, charging electrodes, deflection electrodes, high pressure ink supplies, and guttering and re-circulation systems~\cite{wijshoff-2010-physrep}. DOD printing is thus more versatile than CIJ, also in terms of ink usage and droplet deposition strategies, and therefore the method of choice to achieve high printing speeds and high accuracy while printing complex drop deposition patterns~\cite{derby-2010-arms, wijshoff-2010-physrep, book-hoath, lohse-2022-arfm}. 
{\color{red} In DOD printing, the actuation pressure required to eject a droplet is generated either by a piezoelectric actuator (piezoacoustic DOD printing) or by a vapor bubble that is thermally nucleated (thermal inkjet printing)~\cite{lohse-2022-arfm}. In contrast to piezoacoustic inkjet printing, thermal inkjet printing puts constraints on ink properties in terms of its boiling point and resistivity to high temperatures.} 
The ability {\color{red} of piezoacoustic inkjet printing} to print across a wide range of liquid properties, has opened its use to a multitude of application areas beyond graphics printing, including the fabrication of electronic displays~\cite{shimoda-2003-mrsbull}, electronics printing~\cite{sirringhaus-2000-science, majee-2016-carbon, majee-2017-carbon}, tribology~\cite{vanderKruk2019}, and in life sciences~\cite{villar-2013-science, daly-2015-intjpharm, simaite-2016-sensactuatorsb}. Typically, the inks used in inkjet printing are multicomponent, consisting not only of colored pigments and surfactants, but also of a collection of cosolvents with varying volatilities. Hence, selective evaporation of one or more of these components, {\color{red} following the most volatile one~\cite{Sirignano1999}}, is inherent to inkjet printing~\cite{lohse-2020-natrevphys}.

In inkjet printing, selective evaporation is mainly studied in the context of the drying sessile droplet formed after drop impact on the substrate~\cite{sefiane-2014-acsi, kim-2016-prl}. The dominant theme in this line of research has been the `coffee stain effect'~\cite{deegan-1997-nature}, or its suppression through, e.g. control of paper porosity \cite{Dou2012,Pack2015} or via solutal~\cite{talbot-2015-acsami, kim-2016-prl, karpitschka-2017-langmuir, marin-2019-prf, mouat-2020-prl}, thermal~\cite{Ristenpart2007,lee-2018-prappl} or surfactant-induced~\cite{Kim2016,Marin2016,VanGaalen2021} Marangoni effects. Recent studies have also shown that selective evaporation can induce complex physicochemical hydrodynamics in multicomponent systems, leading to remarkable observations such as segregation~\cite{tan-2016-pnas, tan-2017-softmatter, li-2018-prl,Phlavan2021,Mao2020} and crystallization~\cite{mailleur-2018-prl, li-2020-langmuir}. Interestingly, even at small length scales, evaporation-driven segregation can cause gravity-dominated flows~\cite{edwards-2018-prl, li-2019-prl, diddens-2021-jfm}. 

Selective evaporation is also important on the residual droplets of ink typically present on the nozzle plate~\cite{dejong-2007-apl}. Here, preferential evaporation of an ink component can introduce a concentration gradient, resulting in a Marangoni flow~\cite{dejong-2007-apl, beulen-2007-expfluids}. This flow can transport dirt particles towards the nozzle, which may result in bubble entrainment and nozzle failure~\cite{fraters-2019-prappl,Fraters2019,Li2019,Fraters2021Oscillations} -- events that are highly undesirable in inkjet printing.

A third area in inkjet printing where selective evaporation is important is within the print-nozzle itself. During the continuous ejection of droplets, the ink within the inkchannel and its nozzle remain well-mixed. However, in the course of printing a more complex and multicolored pattern---typically using multiple printheads each with thousands of nozzles---a number of nozzles remain idle during certain periods of time. During this idle period, selective evaporation takes place at the nozzle exit~\cite{Swanson1992,Chauvet2010,Bacchin2022}, thereby changing the local composition and properties of the ink, which can lead to major inaccuracies in droplet volume and speed due to a change in surface tension and/or viscosity of the ink. Hence, it is of utmost importance to be able to probe, and understand the effects of selective evaporation of ink at the nozzle exit in order to mitigate disturbing consequences and develop robust inkjet printing.

\begin{figure*}
	\centering
	\includegraphics[width = 1.9\columnwidth]{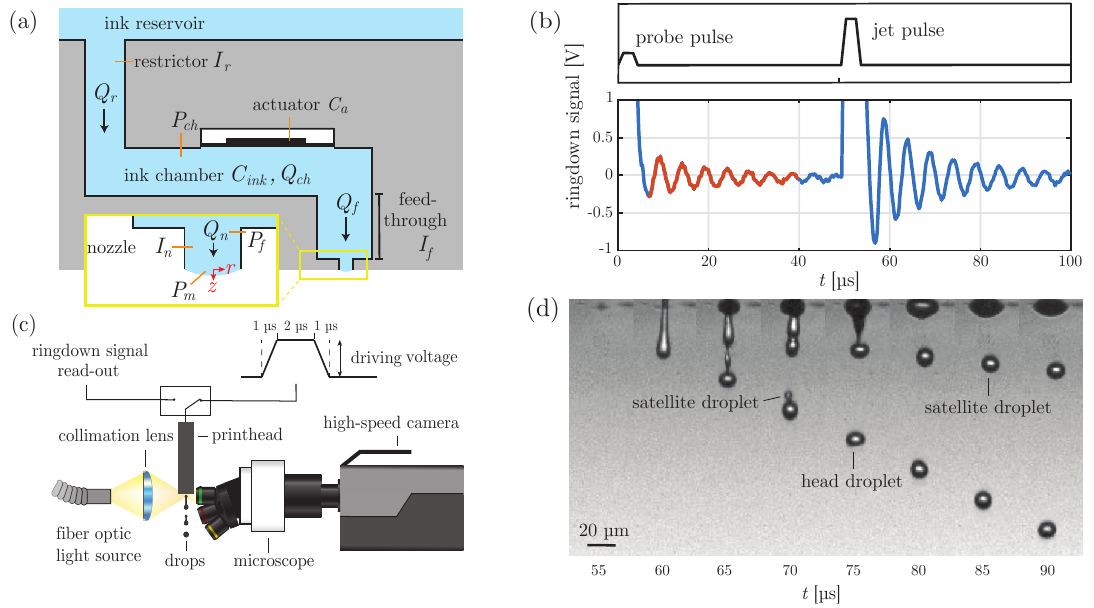}
	\caption{(a) A schematic of the employed ink channel with the ink chamber, restrictor, feedthrough, and nozzle. The ink channel is driven by a piezoelectric actuator attached to a flexible membrane, with compliance $C_a$. Also indicated are the inertances of the restrictor $I_r$, the feedthrough $I_f$, and the nozzle $I_n$. The arrows indicate the direction of the flow rate in the restrictor $Q_r$, the feedthrough $Q_f$, and the nozzle $Q_n$. The pressure in the ink chamber is indicated by $P_{ch}$, the pressure in between the feedthrough and nozzle is $P_f$, and the Laplace pressure at the meniscus is $P_m$. The compliance of the ink $C_{ink}$ in the ink chamber and the flow rate through the ink chamber $Q_{ch}$ are also indicated.
		(b) The driving waveforms and their corresponding piezo signals as recorded by the oscilloscope. The region that is used for the extraction of the ringdown frequency and decay rate is highlighted in red. The actuation pulses are indicated above the graph. 
		(c)  A schematic of the experimental setup. The printhead was driven by a 1-2-\SI{1}{\micro\second} trapezoidal pulse. Subsequently, the piezo connections switched to the read-out circuit to probe the ringdown of the inkchannel acoustics. A light source back-illuminated the jetted droplets and images were captured by a high-speed camera.
		(d) A typical drop formation process with no drying time.
	}
	\label{fig:setup}
\end{figure*}

One way to probe the effect of selective evaporation on the liquid composition within the nozzle is to observe the droplet formation process. However, the extraction of physical liquid properties directly from those observations is non-trivial as there is no direct description of the droplet formation as function of the liquid properties. Instead, it requires a corresponding numerical model that accurately describes the drop formation process~\cite{Sengun1999,Hoath2013}. Another method is to probe the acoustic ringdown characteristics of the inkchannel by measuring the piezo-signal after the piezo has actuated the system~\cite{jeurissen-2009-jasa, wijshoff-2010-physrep}. The acoustics of today's printheads---typically fabricated in silicon using micro-electromechanical systems (MEMS) technology---resemble that of a Helmholtz resonator~\cite{kim-2014-sensactuatorsa,segers-draft}. The acoustic ringdown frequency is then governed by the mass of ink in the nozzle and the restrictor and by the compliance (spring constant) of the ink chamber and its ink volume (Fig.~\ref{fig:setup}(a)). Damping results mainly from viscous dissipation in the restrictor and the nozzle. Therefore, a change in the oscillatory behavior and damping of the ink channel acoustics can directly be connected to changes in ink properties in the nozzle. 

In the present work, we investigate the effects of the idle, or drying time, on the selective evaporation of a multicomponent ink at the nozzle exit of a piezo-acoustic inkjet printhead. We measure the acoustic ringdown signal of the inkchannel, and develop an analytical lumped-element model to gain physical insight into the physicochemical composition of the ink within the nozzle. We also compare the conclusions drawn from the experiments to the results of the numerical simulations of the complete evaporation and jetting process.

The paper is organized as follows: In section II, we describe the experimental procedure and in section III we introduce the analytical model for analyzing the printhead acoustics. In section IV, we present the numerical method and section V contains the results and the corresponding discussion. The paper ends with conclusions and an outlook.

\section{Printhead, model ink, and experimental procedure}

A schematic of the employed inkchannel, which is part of an experimental printhead (Canon Production Printing, Venlo, The Netherlands) is shown in Fig.~\ref{fig:setup}(a). The ink reservoir feeds the ink chamber with its piezo actuator via a restrictor channel. The nozzle with a radius of \SI{8}{\micro\meter} is connected to the ink chamber via a feedthrough channel. The symbols in Fig.~\ref{fig:setup}(a) will be explained in the modeling section. The printhead was driven by a trapezoidal pulse with ramp times of \SI{1}{\micro\second}, and a high time of \SI{2}{\micro\second}. The waveform was generated by an arbitrary waveform generator (Agilent 33220A) and amplified by a broadband amplifier (Falco System WMA-300), resulting in a pull-push motion of the liquid in the nozzle. After driving the piezo, the ringdown of the inkchannel acoustics was measured by switching the piezo-connections to an oscilloscope (Tektronix TDS5034B) via a transimpedance amplifier as described in~\cite{dejong-2007-apl}.

The model ink used in the present study was a mixture of \SI{10}{\wtpercent} glycerol (Sigma-Aldrich) in MilliQ water, which has a density $\rho$ of \SI{1020.4}{\kilo\gram/\meter\cubed}~\cite{Volk2018}, a viscosity $\mu$ of \SI{1.1}{\milli\pascal.\second}~\cite{Cheng2008}, and a surface tension $\gamma$ of \SI{68.3}{\milli\newton/\meter}~\cite{Takamura2012}. To prevent the liquid from dripping out of the printhead due to gravity, the ink channel was always kept at an underpressure of \SI{8}{\milli\bar}. 

The experimental measurement procedure was as follows. First, 999 droplets were jetted at a DOD frequency of \SI{1}{\kilo\hertz} to make sure that the liquid mixture in the nozzle had the same composition as that in the bulk. Whether or not this amount of jetted droplets was enough was verified using the measured resonance frequency, as we will also describe in section~\ref{subsec:recovery}. Next, the jetting was stopped for the predetermined drying time, which led to evaporation from the stationary liquid meniscus at the nozzle exit. The relative humidity in the lab was 38~$\%$ $\pm$ 3~$\%$. By purging nitrogen gas across the nozzle plate, measurements at low relative humidty (0~$\%$) were performed. When the desired drying time had passed (the control parameter in the experiments), the piezo was driven by two pulses: A \SI{5}{\volt} probe pulse and a \SI{20}{\volt} jet pulse (see Fig.~\ref{fig:setup}b). The amplitude of the probe pulse was one fourth of the amplitude of the jet pulse to drive the inkchannel acoustics while preventing liquid to be jetted outward. The second pulse was at full amplitude and produced a droplet. The ringdown signals of both pulses were recorded and the signal due to the probe pulse (red curve in Fig.~\ref{fig:setup}b) was used for further analysis in Matlab, where the ringdown frequency and decay rate of the ringdown signal were determined by fitting a damped cosine: $e^{-\beta t}\cos{\omega t}$, where $\beta$ is the decay rate due to damping, and $\omega$ the angular frequency.  

The droplet formation process driven by the jet pulse was recorded using a high-speed camera (Shimadzu HPV-X2, $10^6$ frames-per-second). It allowed validation of our finite element numerical simulations to thereby gain insight in the local concentration gradients within the nozzle and the feedthrough. The imaging setup (Fig.~\ref{fig:setup}(c)) consisted of a modular microscope (BXFM-F, BXFM-ILHS, Olympus) and a 20$\times$ objective (SLMPLN, Olympus). The resulting imaging resolution was \SI{1.86}{\micro\meter/pixel}. Back-illumination was provided by a fiber-optic light source (LS-M352, Sumita). The waveform generator and camera were triggered with their appropriate delays at nanosecond precision using a pulse-delay generator (BNC 575, Berkeley Nucleonics Corp). A typical droplet formation process is shown in Fig.~\ref{fig:setup}d. The figure shows a liquid jet being ejected, which eventually breaks up to form a head droplet and a satellite droplet~\cite{Fraters2020}.

\section{Analytical model of the printhead acoustics}\label{sec:model}
\begin{figure}
	\centering
	\includegraphics[width=\columnwidth]{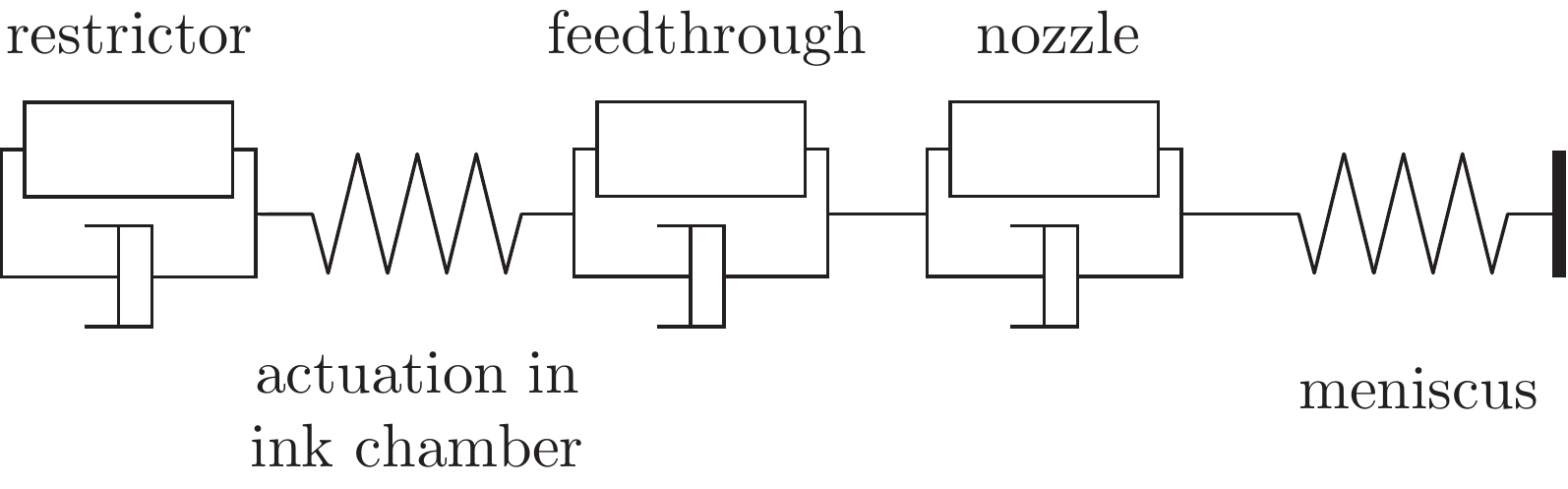}
	\caption{Lumped element model of the ink channel shown in Fig.~\ref{fig:setup}a where the restrictor, feedthrough, and nozzle are composed of a mass and a damper. The ink chamber and meniscus are both represented by a spring.}
	\label{fig:mass}
\end{figure}
A lumped element model, similar to that of ref.~\cite{segers-draft}, was developed to describe the acoustics in the inkchannel to relate the measured ringdown signal to the mixture composition in the nozzle (see Fig.~\ref{fig:mass}). Unlike in~\cite{segers-draft}, we describe the acoustics in the frequency domain to include viscous damping, similar to what was done in~\cite{jeurissen-2009-jasa}. The flow in the nozzle in response to the probe pulse is oscillatory and the resulting flow profile depends on the competition between the inertia of the oscillating velocity field and viscosity, characterized by the Womersley number~\cite{Womersley1955}:
\begin{equation} 
Wo = R_n \sqrt{ \frac{\omega \rho_n}{\mu_n}},
\end{equation}
with $R_n$ the radius of the nozzle, $\omega$ the angular frequency of oscillations, and $\rho_n$ and $\mu_n$ the ink density and dynamic viscosity in the nozzle, respectively. The flow rate $Q_n$ in a cylindrical nozzle is determined by, next to the pressure pulse, the oscillation frequency, and has been derived analytically in closed form by Womersley~\cite{Womersley1955}. Here, it is rewritten as the ratio of the pressure difference across the inertance of the nozzle ($P_f-P_m$, see Fig.~\ref{fig:setup}a) to the flow rate in the nozzle ($Q_n$) to obtain the acoustic impedance $Z_n$ of the nozzle, given by:
\begin{equation} \label{Eq:Zn}
Z_n = \frac{P_{f} - P_m}{Q_n} = i \omega I_n \left(  1 - \frac{2 J_1(i^{3/2}Wo) }{i^{3/2}Wo J_0(i^{3/2}Wo) }	 \right)^{-1},
\end{equation}
with $P_{f}$ the pressure at the nozzle inlet, $P_{m}$ the Laplace pressure at the meniscus (described in more detail after Eq.~\ref{eq:chamber}), and $J_0$ and $J_1$ the ordinary Bessel functions of the first kind of zeroth and first order, respectively. $I_n$ is the acoustic inertance of the nozzle, given by:
\begin{equation}\label{eq:inertance}
I_n = \frac{\rho_n L_n}{A_n},
\end{equation}
with $A_n$ the cross-sectional area of the nozzle and $L_n$ the length of the nozzle. The effective inertance of the nozzle is larger than that given by Eq.~\ref{eq:inertance} as the fluid just outside the nozzle in the feedthrough takes part in the oscillations. This increase in inertance is captured by an increase in nozzle length of: $\Delta L = \pi R_n/4$, as described by Landau and Lifshitz~\cite{Lifshitz1984}, amounting to $I_n$~=~1.08$\times$10$^8$~kg/m$^4$ for the \SI{10}{\wtpercent} glycerol solution. For the acoustic impedance of the restrictor $Z_r$ and the feedthrough $Z_f$, similar equations are used:
\begin{align}
    Z_r &= -\frac{P_{ch}}{Q_r} = i \omega I_r \left(  1 - \frac{2 J_1(i^{3/2}Wo) }{i^{3/2}Wo J_0(i^{3/2}Wo) }	 \right)^{-1},\\
    Z_{f} &= \frac{P_{ch}-P_f}{Q_f} = i \omega I_f \left(  1 - \frac{2 J_1(i^{3/2}Wo) }{i^{3/2}Wo J_0(i^{3/2}Wo) }	 \right)^{-1},
\end{align}
with the corresponding inertance of the restrictor $I_r$~=~1.19$\times$10$^8$~kg/m$^4$ and that of the feedthrough $I_f$~=~1.26$\times$10$^7$~kg/m$^4$. The inner dimensions of the printhead can deviate from the intended values due to fabrication inaccuracies of the MEMS chip. As the exact dimensions could not be measured, we used the length of the restrictor as a fitting parameter.

The total compliance  of the ink channel $C_{ch}$ (volume change per unit pressure) has contributions from the flexible piezo actuator $C_a$ and the volume of ink $V_{ink}$ in the ink chamber ($C_{ink} = V_{ink}\rho c^2$, with $c$ the speed of sound in the ink) which leads to $C_{ch} = C_{a}+C_{ink}$. The total compliance of the employed ink channel was $C_{ch}$~=~10.8$\times$10$^{-21}$~m$^3$/Pa, as obtained from investigations on the resonance behavior of the printhead~\cite{wijshoff-2010-physrep}. The pressure change in the ink chamber (Fig.~\ref{fig:setup}a) due to its change in volume is given by:
\begin{equation} 
P_{ch} = \frac{1}{C_{ch}}(V_r-V_f),
\end{equation}
with $V_r$ the displaced volume from the restrictor into the ink chamber  and $V_f$ that from the ink chamber into the feedthrough. The flow rate into the ink chamber equals $Q_{ch} = Q_r - Q_f$, where the flow rate is the derivative of the displaced volume with respect to time. By assuming simple harmonic motion ($V(t)~=~V_A e^{i \omega t}$, with $V_A$ a constant amplitude), it follows that the acoustic impedance of the compliant ink channel is given by:
\begin{equation} \label{eq:chamber}
Z_{ch} = \frac{P_{ch}}{Q_{ch}} = \frac{1}{i \omega C_{ch}}.
\end{equation}

The surface tension of the meniscus at the nozzle exit adds another compliance to the system. The compliance of the meniscus can be estimated from the amount of liquid volume sustained by the surface tension of the meniscus, as shown in~\cite{kim-2014-sensactuatorsa}. However, instead of assuming that the meniscus protrudes from the nozzle as a hemisphere, which is non-linear, we assume small displacements, for which the meniscus is approximately a paraboloid. The meniscus surface protrudes from the plane of the nozzle plate by a distance $z(r)$, which is a function of the radial coordinate $r$:
\begin{equation}
    z(r) = z_m \left(1 - \frac{r^2}{R_n^2} \right),
\end{equation}
with $z_m$ the maximum height of the meniscus. The Laplace pressure $P_m$ is the product of surface tension $\gamma$ and the surface curvature $\kappa = -\nabla^2 z = 4z_m/R_n^2$. The volume displacement is obtained by integrating the meniscus displacement over the area of the meniscus, resulting in the compliance of the meniscus:
\begin{equation}
    C_m = \frac{V_{m}}{P_{m}} = \frac{z_m \frac{1}{2} \pi R_n^2}{\gamma 4 z_m R_n^{-2}} = \frac{\pi R_n^4}{8 \gamma},
\end{equation}
which is independent of $z_m$ and results in $C_m$~=~23.6$\times$10$^{-21}$~m$^3$/Pa. The acoustic impedance of the meniscus is then given by:
\begin{equation}
    Z_m = \frac{P_m}{Q_n} = \frac{1}{i \omega C_m}.
\end{equation}
 
\begin{figure}
    \centering
    \includegraphics[width=\columnwidth]{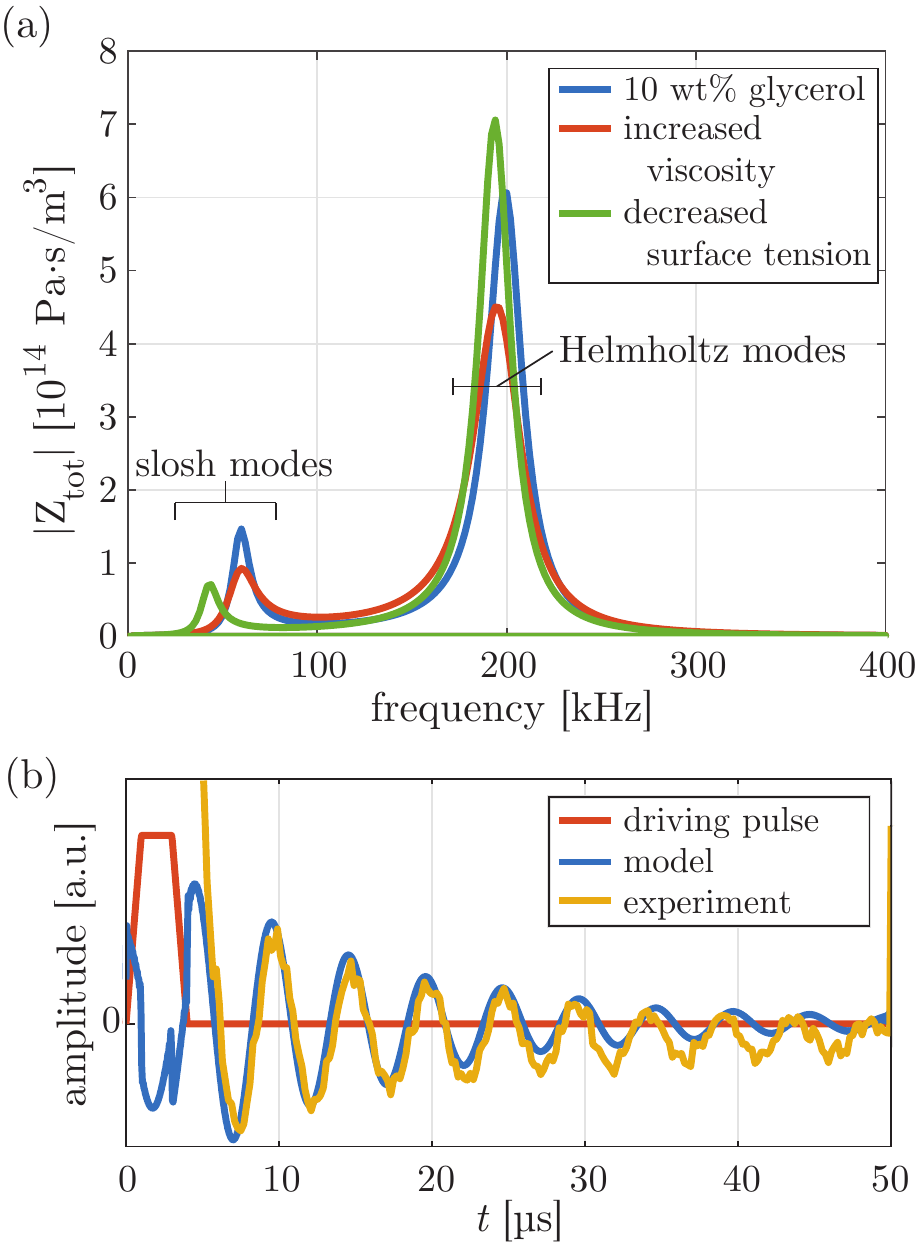}
    \caption{(a) Real part of the total impedance given by Eq.~\ref{eq:ztot} with a \SI{10}{\wtpercent} glycerol concentration in all components of the printhead (blue), with double the viscosity in the nozzle (red), and half the surface tension in the nozzle (green). (b) Driving pulse (red) with the corresponding ringdown signal for the \SI{10}{\wtpercent} glycerol experiment (yellow) with no drying time, and the analytical model (blue).}
    \label{fig:impedance}
\end{figure}

To find the resonance frequencies of the inkchannel represented by the coupled impedances, mass continuity from the feedthrough to the nozzle, i.e. $Q_f = Q_n$, is added to the system of equations in order to balance the number of unknown variables with the number of equations such that the coupled system of equations can be solved. The expressions for the impedances are then written in matrix form~\cite{segers-draft}, as follows: 
\begin{equation}\label{Eq:matrix}
\begin{split}
\begin{bmatrix}
0 & 0 & 0 & 0 & 1 & -1\\
0 & 1 & 0 & 0 & -\frac{1}{i \omega C_m} & 0\\
0 & -1 & 1 & 0 & {\scriptstyle -i\omega I_n f(Wo_n)} & 0\\
1 & 0 & -1 & 0 & 0 & {\scriptstyle -i\omega I_f f(Wo_f)}\\
1 & 0 & 0 & -\frac{1}{i\omega C_{ch}} & 0 & \frac{1}{i\omega C_{ch}}\\
-1 & 0 & 0 & {\scriptstyle -i\omega I_r f(Wo_r)} & 0 & 0
\end{bmatrix} \\
\times
\hfill
\begin{bmatrix}
P_{ch}\\
P_m\\
P_f\\
Q_r\\
Q_n\\
Q_f
\end{bmatrix}
= 0.
\end{split}
\end{equation}
The corresponding determinant is found to be:
\begin{equation}\label{Eq:det}
\begin{split}
        &-\omega^2 I_r f(Wo_r)(I_f f(Wo_f)+I_n f(Wo_n))\\
        &+\frac{I_n f(Wo_n)+I_f f(Wo_f)+I_r f(Wo_r)}{C_{ch}}\\
        &+ \frac{I_r f(Wo_r)}{C_m}-\frac{1}{\omega^2C_{ch}C_m},
\end{split}
\end{equation}
with
\begin{equation}
f(Wo)=\left(  1 - \frac{2 J_1(i^{3/2}Wo) }{i^{3/2}Wo J_0(i^{3/2}Wo) }	 \right)^{-1}. 
\end{equation}
The determinant in Eq.~\ref{Eq:det} of the $6 \times 6$ matrix in Eq.~\ref{Eq:matrix} is equated to zero to find the resonance frequencies of the system:
\begin{equation} \label{det}
\begin{split}
&-\omega^4 I_r f(Wo_r) \left[ I_f f(Wo_f)+I_n f(Wo_n) \right] \\
&+\omega^2 \left[ \frac{I_n f(Wo_n)+I_f f(Wo_f)+I_r f(Wo_r)}{C_{ch}} + \frac{I_r f(Wo_r)}{C_m} \right] \\
& - \frac{1}{C_{ch}C_m} = 0.
\end{split}
\end{equation}

The complex solution of this determinant gives the oscillation frequencies of the system, where the real part is the ringdown frequency. For a finite viscosity, the eigenvalue is complex and the positive imaginary part is the decay rate, showing that energy is dissipated by damping in the system. As the equation is not a polynomial function, we employed Matlab's fsolve function (Levenberg-Marquardt method) to find the roots. The root corresponding to the Helmholtz resonance mode of the system was found by providing the Helmholtz resonance frequency of the inviscid case as the initial value. First, this was done for the case where the model ink viscosity, density, and surface tension correspond to a \SI{10}{\wtpercent} glycerol in water mixture. The root that was found was then used for the next iteration of the model where the concentration of glycerol in the nozzle was increased by \SI{1}{\wtpercent}. This process was repeated until the \SI{100}{\wtpercent} glycerol concentration was reached. The viscosity, density, and surface tension of glycerol-water mixtures were obtained from~\cite{Cheng2008,Volk2018,Takamura2012}. The glycerol concentration in the analytical model is only changed in the nozzle and it was kept constant in the rest of the system at all times.

In addition to the ringdown frequency and decay rate, we also compute the ringdown signal itself. We therefore first determine the total lumped acoustic impedance of the ink channel:
\begin{equation}\label{eq:ztot}
Z_{tot} = \left(\frac{1}{Z_r} + \frac{1}{Z_n+Z_f+Z_m} +\frac{1}{Z_{ch}} \right)^{-1}.
\end{equation}
The flow rates through the feedthrough and nozzle toward the meniscus are the same, assuming incompressibility, and therefore, $Z_n$, $Z_f$, and $Z_m$ are summed. The inverse sum is taken with the other components as these impedances are connected in parallel. The real part of $Z_{tot}$ is plotted in Fig.~\ref{fig:impedance}a as function of frequency for \SI{10}{\wtpercent} glycerol (solid blue curve), for \SI{10}{\wtpercent} glycerol with its viscosity artificially doubled (red curve), and \SI{10}{\wtpercent} glycerol with its surface tension artificially halved (green curve). The peak at \SI{200}{\kilo\hertz} corresponds to the Helmholtz resonance mode of the system where the masses (inertances) of the restrictor and nozzle oscillate $180^\circ$ out-of-phase and the spring constant is given by the compliance of the ink chamber~\cite{segers-draft}. The low frequency mode around \SI{60}{\kilo\hertz} corresponds to the slosh mode that is characterized by the in-phase movement of the masses of the nozzle, feedthrough, ink chamber, and restrictor against the compliance of the meniscus~\cite{Dijksman2018} such that for the inviscid case:
\begin{equation}
f_{slosh} = \frac{1}{2 \pi}\sqrt{\frac{1}{C_m(I_r + I_f +  I_n)} } = \SI{69}{\kilo\hertz}.
\end{equation}
Note that the inertance of the ink chamber was neglected as it is much smaller than the other contributions. Also note from Fig.~\ref{fig:impedance}a that both the frequency of the modes and their damping (width of the peak) are sensitive to the liquid properties in the nozzle. The frequency of the slosh mode is mostly affected by a change in surface tension, while its damping (peak width) is only sensitive to viscosity. The Helmholtz resonance frequency decreases by an increase in viscosity and a decrease in surface tension. The damping of the Helmholtz mode increases with an increase in viscosity.\\

The computation of the ringdown signal is continued by relating the pressure in the channel to the piezo driving voltage. In the linear regime of the piezo, the volume change at constant pressure in the ink channel is proportional to the driving voltage with a proportionality constant $\alpha$ (units of m$^3$/V). Using the definition of the acoustic impedance $Z=P/Q$, the pressure in the ink channel in response to a driving pulse can be expressed as:
\begin{equation}
p = i \omega \alpha Z_{tot} V_{pulse}(\omega),
\end{equation}
with $V_{pulse}(\omega)$ the discrete Fourier transform of the time-dependent driving voltage.

The piezo ringdown signal in the experiments is a measure of the piezo current~\cite{Groninger2008}. The linearized relation between piezo current and pressure in the ink channel is given by~\cite{jeurissen-2009-jasa}:
\begin{equation}
I(\omega) = i \omega \alpha p,
\end{equation}
such that the modeled piezo ringdown signal in the frequency domain becomes:
\begin{equation} \label{Eq:Iomega}
I(\omega) = - \omega^2 \alpha^2 Z_{tot} V_{pulse}(\omega).
\end{equation}
Eq.~(\ref{Eq:Iomega}) is then inverse-Fourier-transformed to obtain the piezo ringdown signal in the time domain. A comparison between the modeled piezo ringdown signal and the experimental ringdown signal is shown in Fig.~\ref{fig:impedance}b. The experimental ringdown signal only appears after 5~$\upmu$s because the amplitude of the probe pulse is a lot higher than the amplitude of the ringdown signal (see Fig.~\ref{fig:setup}b). Figure~\ref{fig:impedance}b shows the modeled piezo ringdown signal for \SI{10}{\wtpercent} glycerol, along with the experimental result for a drying time of \SI{1.5}{\milli\second}. The comparison between the results from the model and from the experiment show good agreement.

\section{Numerical model}\label{sec:num}

While the average concentration of glycerol in the nozzle can be estimated from the ringdown signal using the analytical model, a numerical model can be used to simulate the drying process and thereby provide key insight into the dynamical process of glycerol diffusion leading to its enhanced concentration. Moreover, the simulations can provide insight into the local distribution of glycerol in the nozzle and the jetted droplet, information that is not available in the experiment. 


Simulating the entire process, i.e. the selective evaporation of water from the nozzle followed by a jetting event, is a challenging problem. The first challange is that the time scales are drastically different, namely evaporation happening for a duration of hundreds of seconds, whereas the jetting is on the order of microseconds. This demands a flexible and stable temporal integration method that can easily switch the relevant time scales by several orders of magnitude. Furthermore, the scenario is inherently a multi-component and multi-phase problem with phase transitions and mass transfer, while the fluid properties depend on the local liquid composition. Consequently, the numerical implementation requires to allow for multi-component mass transfer across the liquid-gas interface, account for Marangoni flow, and must consider local variations of the mass density and viscosity. Finally, resolving the entire jetting dynamics would demand considering the fluid-structure interaction of the actuating piezo and the coupling between the free-surface flow dynamics at the nozzle and acoustics in the chamber. In this section, we describe how these challenges are tackled in the present simulation framework.

As a general framework, we have used a sharp-interface arbitrary Lagrangian-Eulerian finite element method (ALE-FEM) expressed in axisymmetric cylindrical coordinates. This comes with the benefit that the liquid-gas interfaces are always exactly represented by sharp curves, which easily allows to incorporate Marangoni flow and mass transfer. Furthermore, FEM is solved implicitly via Newton's method, which provides a stable solution method along with flexible time stepping on the two different time scales of evaporation and jetting.
The implementation is based on the finite element library \textsc{oomph-lib}~\cite{oomphlib,Heil2006}.

\subsection{Evaporation phase}
During the evaporation phase, the fluid dynamical equations in both the surrounding gas and the liquid phase are solved. Since there is no actuation, i.e. no applied pulse, all parts of the driving are deactivated in this phase. However, if long drying times are considered, it is important to consider the entire system, i.e. from the nozzle over the feedthrough, the chamber and the restrictor into the ink reservoir domain. Otherwise, the diffusive replenishment of water from the ink reservoir domain is not accurately accounted for in the long-time limit. In the following, the governing equations for solving the evaporation phase are described. These resemble the equations which have been successfully used in previous works on the evaporation of multi-component droplets on substrates, e.g. in~\cite{Li2020}.

\begin{figure}[ht]
\centering
\includegraphics[width=.7\columnwidth]{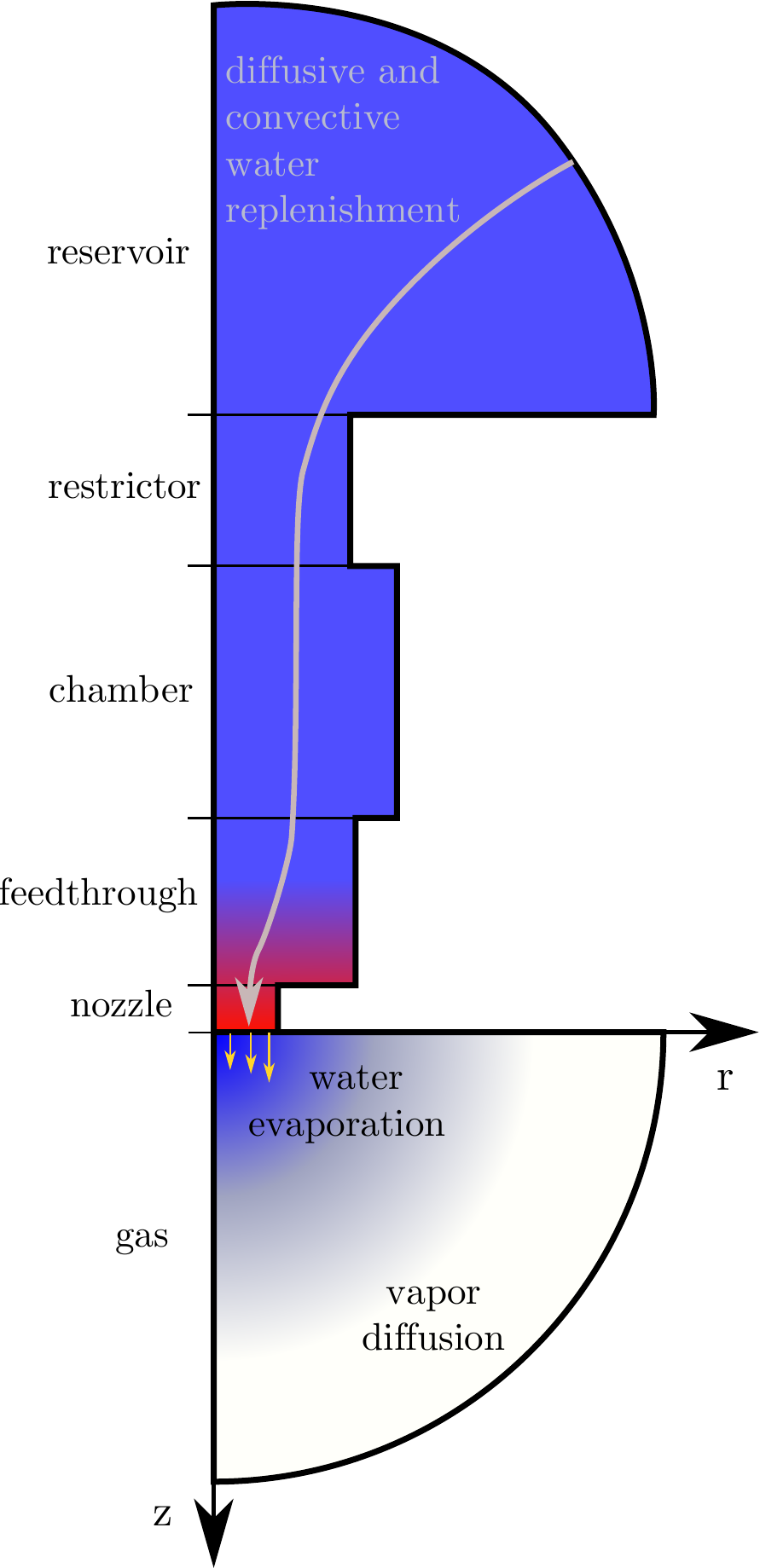}
\label{fig:FEM:evapscheme}
\caption{Schematic of the evaporation phase (not true to scale). The printhead geometry is assumed to be axisymmetric. Water evaporates into the gas phase, leaving behind an enhanced glycerol concentration. Water is replenished by diffusion and convection through the entire system towards the nozzle, as indicated by the gray arrow.}
\end{figure}

\subsubsection{Vapor diffusion in the gas phase}
We assume diffusion-limited evaporation, i.e. the evaporation rate of water can be obtained by solving the vapor diffusion equation for the partial mass density $\VapConcW$ of water vapor in the gas phase by:
\begin{align}
\partial_t\VapConcW= D_\text{vap} \nabla^2 \VapConcW,
\end{align}
subject to the boundary conditions:
\begin{align}
\VapConcW&=\VapConcWVLE(\MassFracW) &\text{at the liquid-gas interface}\\
\VapConcW&=\VapConcWInfty & \text{far away}\,,
\end{align}
i.e. the vapor-liquid equilibrium (VLE) according to Raoult's law $\VapConcWVLE(\MassFracW)$ at the liquid-gas interface, where $\MassFracW$ is the concentration of water (as weight fraction) in the liquid. For this condition, the liquid water concentration $\MassFracW$ is converted to mole fractions, and the corresponding activity coefficient for the glycerol-water mixture is calculated via the group contribution method AIOMFAC~\cite{Zuend2011} \footnote{http://www.aiomfac.caltech.edu/} to account for the non-ideality of the mixture within Raoult's law.
The ambient water vapor concentration $\VapConcWInfty$ far away is not directly imposed at the distant boundaries of the considered gas domain, since it would induce considerable errors originating from the finite size of the considered gas mesh. Instead a Robin boundary condition mimicking an infinite domain is used, which is based on a multi-pole expansion truncated at monopole order~\cite{Diddens2017}.
The mass-transfer rate of water is given by the diffusive flux at the liquid-gas interface, i.e. by taking the normal derivative:
\begin{align}
\MassTransW=-D_\text{vap}\partial_n\VapConcW.
\end{align}
Advective transport, e.g. due to Stefan flow, is irrelevant for water at room temperature \cite{Carle2016,DiddensJFM2017}. In comparison to water, the volatility of glycerol is negligible, so that we do not account for glycerol evaporation here.

\subsubsection{Multi-component flow in the liquid phase}

The bulk flow in the liquid phase is governed by the Navier-Stokes equations with a composition-dependent mass density $\rho$ and viscosity $\mu$ together with the advection-diffusion equation for the water concentration $\MassFracW$ with a composition-dependent diffusivity $D(\MassFracW)$,
\begin{align}
\rho\left(\partial_t\mathbf{u}+\mathbf{u}\cdot\nabla\mathbf{u}\right)&=-\nabla p +\nabla\cdot\left[\mu\left(\nabla\mathbf{u}+(\nabla\mathbf{u})^\mathrm{t}\right)\right] \label{eq:FEM:navstokes},\\
\partial_t \rho + \nabla\cdot\left(\rho\mathbf{u}\right)&=0 \label{eq:FEM:contieq},\\
\rho\left(\partial_t \MassFracW + \mathbf{u}\cdot\nabla \MassFracW\right)&=\nabla\cdot\left(\rho D\nabla\MassFracW\right). \label{eq:FEM:compoadvdiff}
\end{align}
The composition-dependent properties, i.e. $\rho(\MassFracW)$, $\mu(\MassFracW)$, $D(\MassFracW)$ and also the surface tension $\sigma(\MassFracW)$ were obtained by fitting experimental data from refs. \cite{Takamura2012,DErrico2004,Cheng2008}.
At the liquid-gas interface, normal and tangential stress balances, i.e. Laplace pressure and Marangoni shear, are applied without consideration of the stresses in the gas phase, which can be disregarded due to the small density and viscosity ratios:
\begin{align}
\mathbf{n}\cdot\mathbf{T}\cdot\mathbf{n}&=\kappa\sigma \label{eq:FEM:laplpress}\\ 
\mathbf{n}\cdot\mathbf{T}\cdot\mathbf{t}&=\nabla_\text{S}\sigma \label{eq:FEM:marastress},
\end{align}
with the stress tensor $\mathbf{T}=-p\mathbf{1}+\mu(\nabla\mathbf{u}+(\nabla\mathbf{u})^\mathrm{t})$ and the normal and tangent $\mathbf{n}$ and $\mathbf{t}$, respectively. $\kappa$ is, as before, the curvature of the interface and $\nabla_\text{S}$ is the surface gradient operator.
The kinematic boundary condition considering water evaporation reads
\begin{align}
\rho\left(\mathbf{u}-\mathbf{u}_\text{I}\right)\cdot\mathbf{n}=\MassTransW  \label{eq:FEM:kinbc},
\end{align}
which connects the normal liquid bulk velocity $\mathbf{u}$ with the normal interface velocity $\mathbf{u}_\text{I}$ via the evaporation rate $\MassTransW$. The liquid-solid interfaces within the simulated printhead geometry are no-slip boundary conditions.
Finally, the evaporation of water leads to a change of the liquid composition near the interface, which is incorporated via the boundary condition
\begin{align}
-\rho D\nabla\MassFracW\cdot\mathbf{n}=(1-\MassFracW)\MassTransW\, . \label{eq:FEM:evapcompochange}
\end{align}
The far field in the reservoir is again mimicking an infinite domain by a far-field Robin boundary condition. Furthermore, a constant underpressure of \SI{8}{\milli\bar} is applied as in the experiments, which results in a slightly inwardly curved meniscus.

\subsection{Probe pulse and jetting pulse}
When it comes to jetting, the relevant time scales are several orders of magnitude smaller than those during the drying phase. Given the short time scales and the fast convection velocities during jetting, the diffusion-limited evaporation model as used during the evaporation phase is also questionable in this stage. Therefore, the gas phase and with it the evaporation dynamics are disregarded during the jetting process. The friction of the jetted droplet in the gas phase is in general not entirely negligible, but the influence on the drop formation is. This has been shown by the excellent agreement between experiments and frictionless numerics in slender jet approximation~\cite{vdBos2014}.
In the ALE-FEM simulation here, the deactivation of the gas phase and evaporation implies setting $\MassTransW=0$ in Eqs.~\eqref{eq:FEM:kinbc} and \eqref{eq:FEM:evapcompochange}. For the comparison with the analytical model evaporation is not considered at all, but the nozzle is artificially filled with a prescribed glycerol concentration.

\begin{figure}[ht]
\centering
\includegraphics[width=.8\columnwidth]{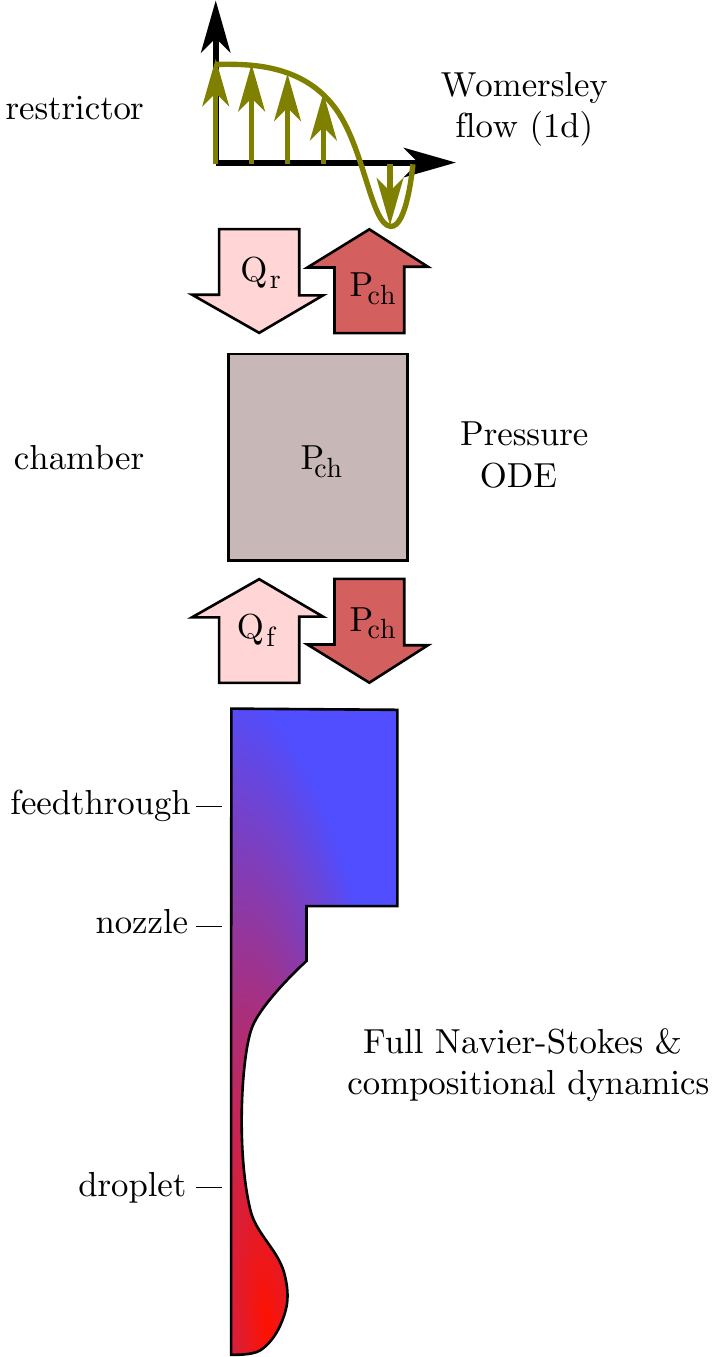}
\label{fig:FEM:jetscheme}
\caption{Schematic during the probe pulse and in the jetting phase (not to scale). The axial restrictor flow is solved on a 1D radial mesh, the chamber pressure is approximated by the ODE of \cref{eq:FEM:chamberode} and only the part comprising feedthrough, nozzle, and the droplet formation is solved by the full flow and compositional equations. The color gradient from blue to red sketches the increase in glycerol concentration.}
\end{figure}

\subsubsection{Modeling of the chamber and the restrictor dynamics}
To account for the acoustics in the chamber, the simulation domain is truncated at the transition from the feedthrough to the chamber. The flow in the chamber is hence not solved directly, but instead the dynamics is characterized by the chamber pressure $\PChamber$ like in the analytical model, Eq.~5. The chamber pressure dynamics is therefore approximated by the following first order ODE:
\begin{align}
\Compliance\dotPChamber=-\VoltToVolumeFactor\dotV_{pulse} + Q_f - Q_r\,. \label{eq:FEM:chamberode}
\end{align}
Here, $\Compliance$ is the acoustic compliance of the chamber, $\VoltToVolumeFactor$ is a conversion factor from the applied voltage to a displaced volume due to the actuation by the pulse $V_{pulse}$. It is noteworthy that $\VoltToVolumeFactor$ is the only free parameter in the entire simulations, which has been fitted to reproduce the best match with the experimental jetting (cf. Fig.~5). Finally, the chamber pressure is influenced by in- and outflow from both sides, i.e. the feedthrough and the restrictor, which are represented by the terms $Q_f$ and $Q_r$, respectively.
The flow in the feedthrough and the nozzle is solved as before by the full multi-component flow dynamics, i.e. \cref{eq:FEM:navstokes,eq:FEM:contieq,eq:FEM:compoadvdiff,eq:FEM:laplpress,eq:FEM:marastress,eq:FEM:kinbc}.
However, at the top of the feedthrough, where it is usually connected to the chamber, the radial velocity $u_r$ is set to zero, whereas the chamber pressure $\PChamber$ (minus the constant underpressure of $\SI{8}{\milli\bar}$) is imposed as driving force. The volume flux $Q_f$ is then directly obtained by integration of the axial velocity over the fictive interface to the chamber at the top of the feedthrough, i.e.
\begin{align}
Q_f=2\pi\int u_z \,r\,\mathrm{d}r \,. \label{eq:FEM:volfluxfeed}
\end{align}
The flow in the restrictor is treated in a similar way. However, since the restrictor geometry can be assumed to be a long cylinder (radius $R_\text{r}$ and length $L_\text{r}$) and given the fact that the liquid mixture inside the restrictor is nearly homogeneous, it is sufficient to solve the axial Womersley flow \cite{Womersley1955} on a one-dimensional radial mesh, i.e.
\begin{align}
\rho\partial_t u_z=\frac{\PChamber}{L_\text{r}}+\mu\left(\frac{1}{r}\partial_r u_z +\partial_r^2 u_z\right)
\end{align}
with $u_z|_{r=R_\text{r}}=0$. The feedback to the chamber pressure $\PChamber$ via the volume flux $Q_r$ is then calculated analogously to \cref{eq:FEM:volfluxfeed}.

\begin{figure*}
	\centering
	\includegraphics[width=\textwidth]{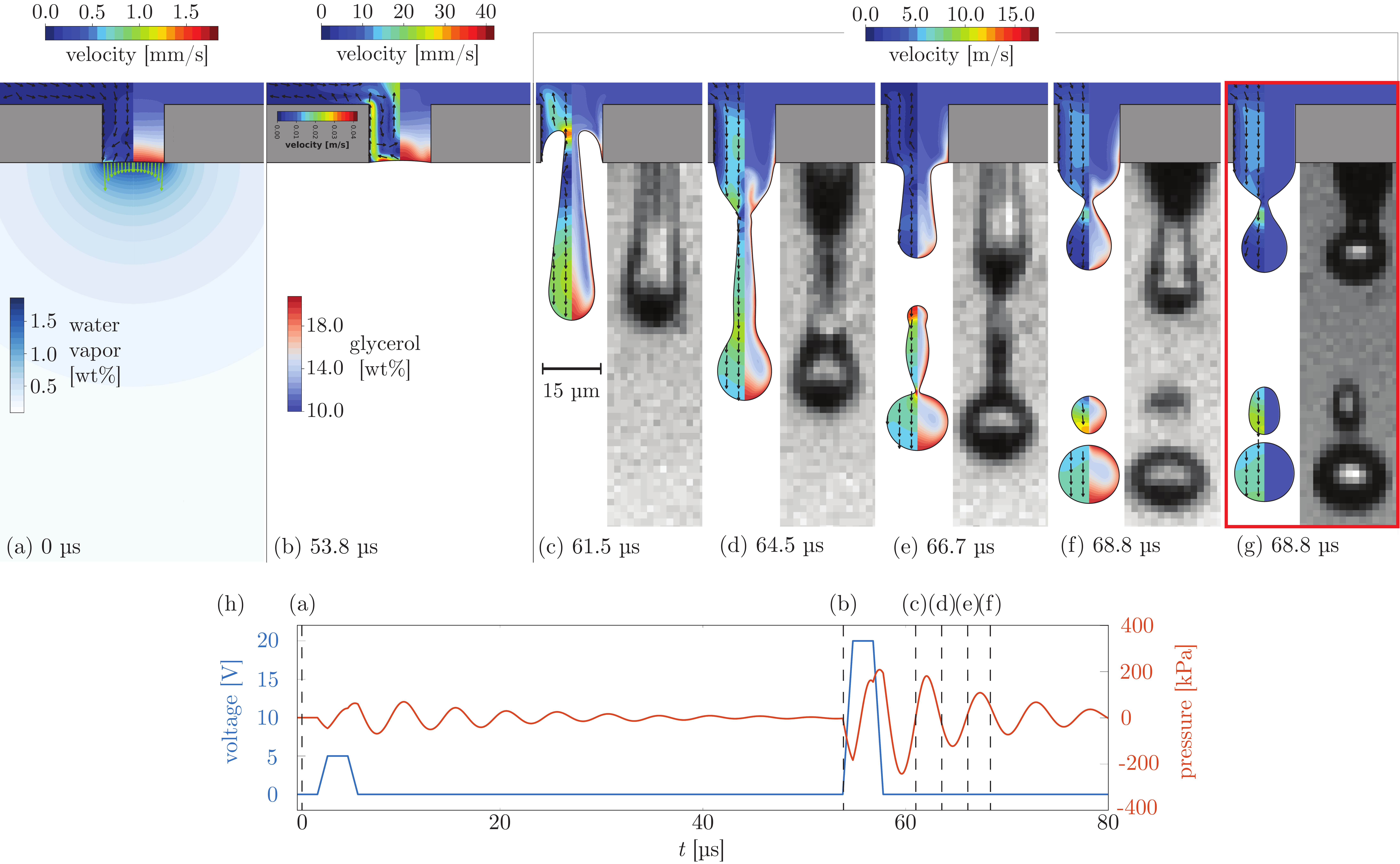}
	\caption{Snapshots from numerics and the corresponding experiment for the droplet formation process with a drying time of \SI{100}{\milli\second} (Video available online). The numerical simulations provide two color schemes inside the liquid: the left hand side of the nozzle shows the velocity profile while the right hand side shows the glycerol concentration. The water evaporation rate is around \SI{100}{\gram/(\meter\squared\second)}.  Snapshots (a) and (b) show the nozzle before droplet formation, and (c)-(f) show the droplet formation process. (g) Snapshot at same time as (f) but for the case of \SI{1.5}{\milli\second} drying time. (h) Demonstrates the actuation pulses (blue) and numerical pressure signal in the ink chamber (red), where the time instants of the snapshots are also indicated.}
	\label{fig:Result_jetting}
\end{figure*}

\subsubsection{Sharp-interface ALE method with topological changes}
While the sharp-interface ALE method has the benefits of easily and accurately incorporating Marangoni flow and evaporation, one of its major drawbacks compared to e.g. volume-of-fluid or phase-field models is the treatment of topological changes, i.e. the pinch-off of droplets from the jet and their potential in-air coalescence.
For simple axisymmetric problems, however, these events can be treated by mesh reconstruction, i.e. after each accepted time step, the liquid-gas interface is tested for parts that run nearly parallel to the axis of symmetry. If these are close to the axis (i.e. within \SI{2}{\percent} of the nozzle radius), the pinch-off position is estimated by finding the thinnest spot of the tail that also shows a profile of local relative outflux, i.e. a relative velocity that changes sign in the vicinity. Whenever such a position can be found, the liquid-gas mesh is artificially split and reconnected to the axis of symmetry. Afterwards, a new separated mesh is constructed and all relevant fields are interpolated from the previous mesh, whereby the data of nodes at the liquid-gas interface are interpolated by the data stemming from the previous, still connected, interface.
A similar treatment is done for droplet coalescence. Here, the liquid-gas interface is re-connected, whenever two distinct parts of the interface overlap at the axis of symmetry.

While this method introduces an artificial length scale, i.e. the thickness threshold for a pinch-off to occur, all other numerical methods already intrinsically have these artificial scales, be it the cell size in a volume-of-fluid approach, the interface thickness in a phase field approach or the regularization radius in the slender jet (lubrication theory) method~\cite{Driessen2011}. The method used here for topological changes also showed perfect agreement with experiments on droplets colliding in mid-air~\cite{Hack2021}.

\section{Results \& Discussion}

\subsection{Influence of drying on drop formation}

Figure~\ref{fig:Result_jetting} compares snapshots from numerical simulations and experiments for a drying time of \SI{100}{\milli\second}. Figure~\ref{fig:Result_jetting}(a) shows the water-vapor concentration field around the nozzle after a drying time of \SI{100}{\milli\second}. The evaporation rate is indicated by the green arrows and it is at maximum at the nozzle boundaries, as expected. The velocity field in the liquid is shown in the left half of each numerical snapshot (colorbar on top of the snapshots). The right half of the numerical snapshots shows the glycerol concentration (colorbar constant for all images). Good agreement is observed between the numerical and the experimental results. Note that the probe pulse alters the glycerol distribution at the nozzle exit. Furthermore, we observe that the glycerol enriched liquid at the nozzle exit forms a shell around the jet and subsequently around the droplets.

The final snapshot in Fig~\ref{fig:Result_jetting}(g) shows the comparison between numerics and experiment for when no drying has taken place (which is taken as \SI{1.5}{\milli\second} in the experiments). Note that the drop formation is very similar to the case with \SI{100}{\milli\second} drying time even though the doubled glycerol concentration at the nozzle exit after \SI{100}{\milli\second} drying time (Fig.~\ref{fig:Result_jetting}(a)). Nevertheless, the modified liquid properties will most likely change the drying and spreading behavior of the printed droplet and thus this observation is in any case important for the overall inkjet printing process.

In Fig.~\ref{fig:Result_jetting}h we see in blue the driving voltage of the piezo, with first the probe pulse and later the jet pulse. As a response to these driving pulses, we see in red the numerically obtained pressure in the ink chamber. The next section focuses on the measurements of this ringdown signal of the ink channel acoustics, to study whether the changes in ink properties in the nozzle due to selective evaporation can be acoustically monitored.

\subsection{Acoustically probing the drying phenomenon}\label{txt:drying}

\begin{figure}
	\centering
	\includegraphics[width=.8\columnwidth]{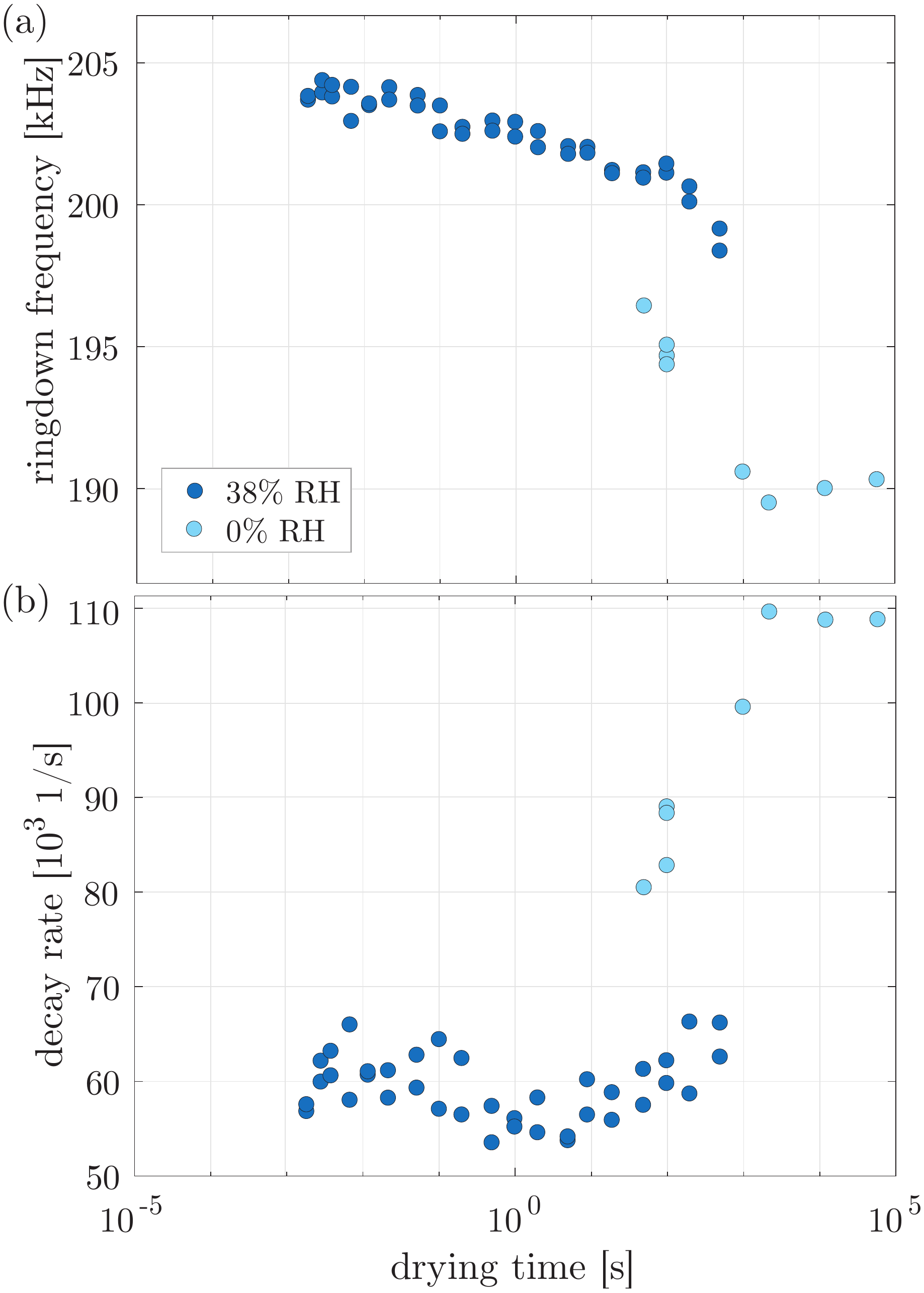}
	\caption{Experimentally measured (a) ringdown frequency and (b) decay rate of the ringdown signal of the probe pulse for different drying times at 38~$\%$ relative humidity (RH) and 0~$\%$ relative humidity (RH).}
	\label{fig:Result_PAINt}
\end{figure}

The experimental results of the ringdown frequency and decay rate for different drying times are shown in Fig.~\ref{fig:Result_PAINt}a and Fig.~\ref{fig:Result_PAINt}b, respectively. For the experiments at a relative humidity of 38~$\%$, there are two measurements points at each drying time, each from a different experiment, demonstrating its reproducibility. From Fig.~\ref{fig:Result_PAINt}a, it can be observed that the ringdown frequency decreases with increasing drying time, and the decay rate in Fig.~\ref{fig:Result_PAINt}b shows an increase with drying time, but only after \SI{1}{\second}. Furthermore, a decrease in humidity seems to amplify the changes in ringdown frequency and decay rate, as observed from the 0~$\%$ relative humidity measurements in Figs.~\ref{fig:Result_PAINt}a and \ref{fig:Result_PAINt}b (light blue datapoints). This suggests that these changes are a consequence of the faster drying process. Finally, the low relative humidity datapoints seem to plateau after a drying time of \SI{2000}{\second} for both the ringdown frequency and the decay rate. This plateau will be explained later from our numerical simulations.

\begin{figure}
	\centering
	\includegraphics[width=\columnwidth]{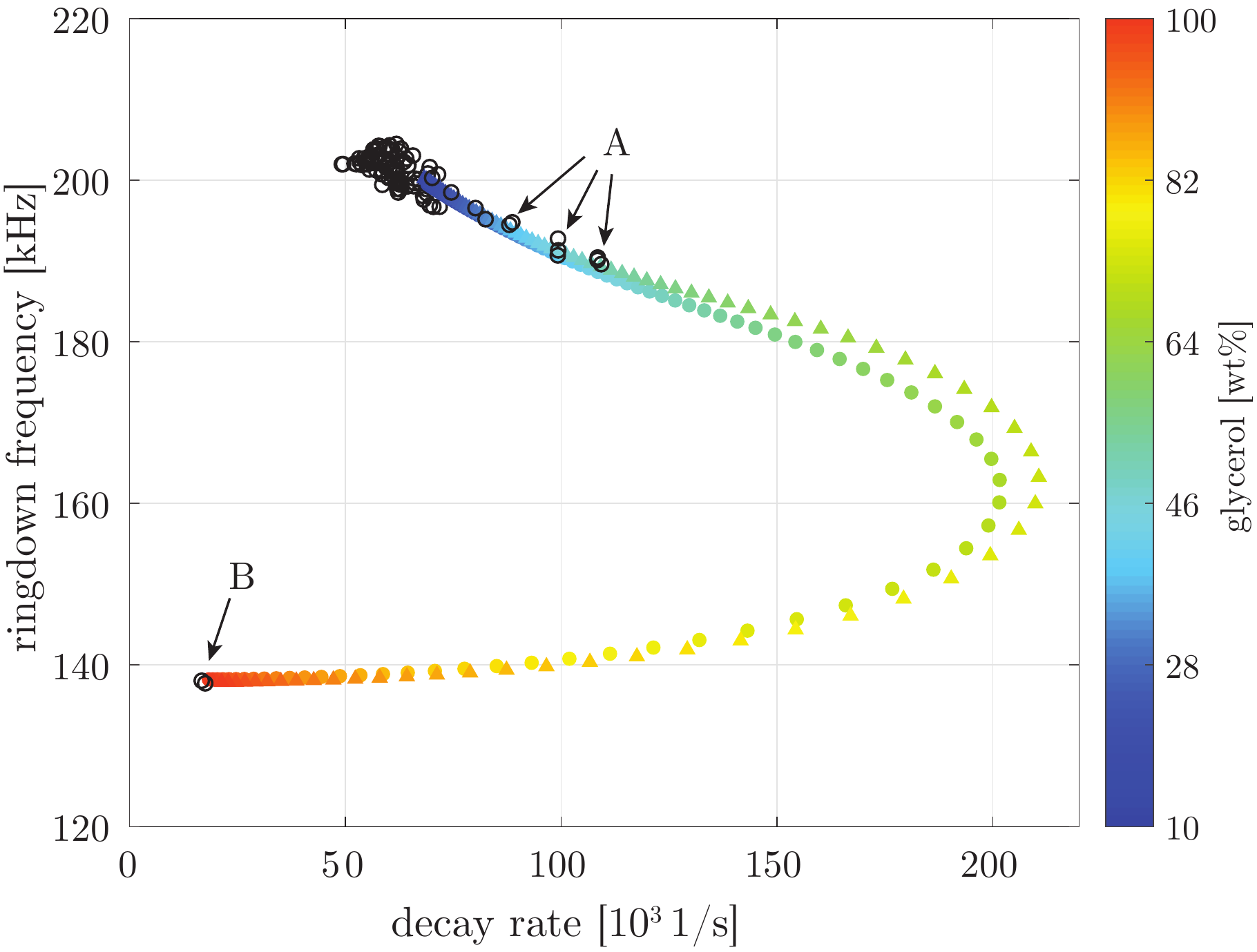}
	\caption{Ringdown frequency and decay rate obtained from experiments (open o), analytical model (filled o), and numerical simulations ($\bigtriangleup$). The colorbar indicates the average glycerol concentration inside the nozzle for the analytical and numerical model. The arrows at A point towards the experimental results obtained at 0$\%$ relative humidity. The arrow at B points toward the results from the manually closed nozzle.}
	\label{fig:damp_vs_freq}
\end{figure}

First, the analytical model described in section~\ref{sec:model} is used to show that the ringdown frequency and decay rate change by varying the glycerol concentration in the nozzle. The resulting output from the model is shown in Fig.~\ref{fig:damp_vs_freq} together with all the experimental results from Fig.~\ref{fig:Result_PAINt}. The experimental results for a humidity of 0$\%$ are indicated by the arrows at A. It can be observed that the decrease in ringdown frequency and increase in decay rate are a result of the increasing glycerol concentration in the nozzle, and that there is a match between the trends of the experiment and the analytical model. Note that the analytical model describes a homogeneous glycerol concentration in the nozzle, while the experiments are expected to contain a gradient in concentration. The arrow at B shows the result for a closed nozzle exit (obtained by manually blocking the nozzle exit). This provides a ringdown frequency and decay rate at the other end of the curve, that also matches very well with the model. The data in Fig.~\ref{fig:Result_PAINt} can be physically explained by describing the printhead as a mass-spring-mass and damper system, with the two moving masses representing the restrictor and the nozzle as they have the highest acoustic inertance. The amplitude of the motion of the mass that represents the nozzle decreases with increasing glycerol concentration due to increased density and viscosity. The point where the glycerol concentration is at its maximum (see Fig.~\ref{fig:damp_vs_freq} ) is the location where the mass of the nozzle does not move anymore, decoupling it from the acoustics inside the printhead and thus effectively rendering a closed nozzle, which is exactly what happens when the nozzle is manually closed off. The good agreement between the analytical model and the experimental measurements demonstrates that compositional changes of the ink in the nozzle can be acoustically measured.

\begin{figure}
	\centering
	\includegraphics[width=.88\columnwidth]{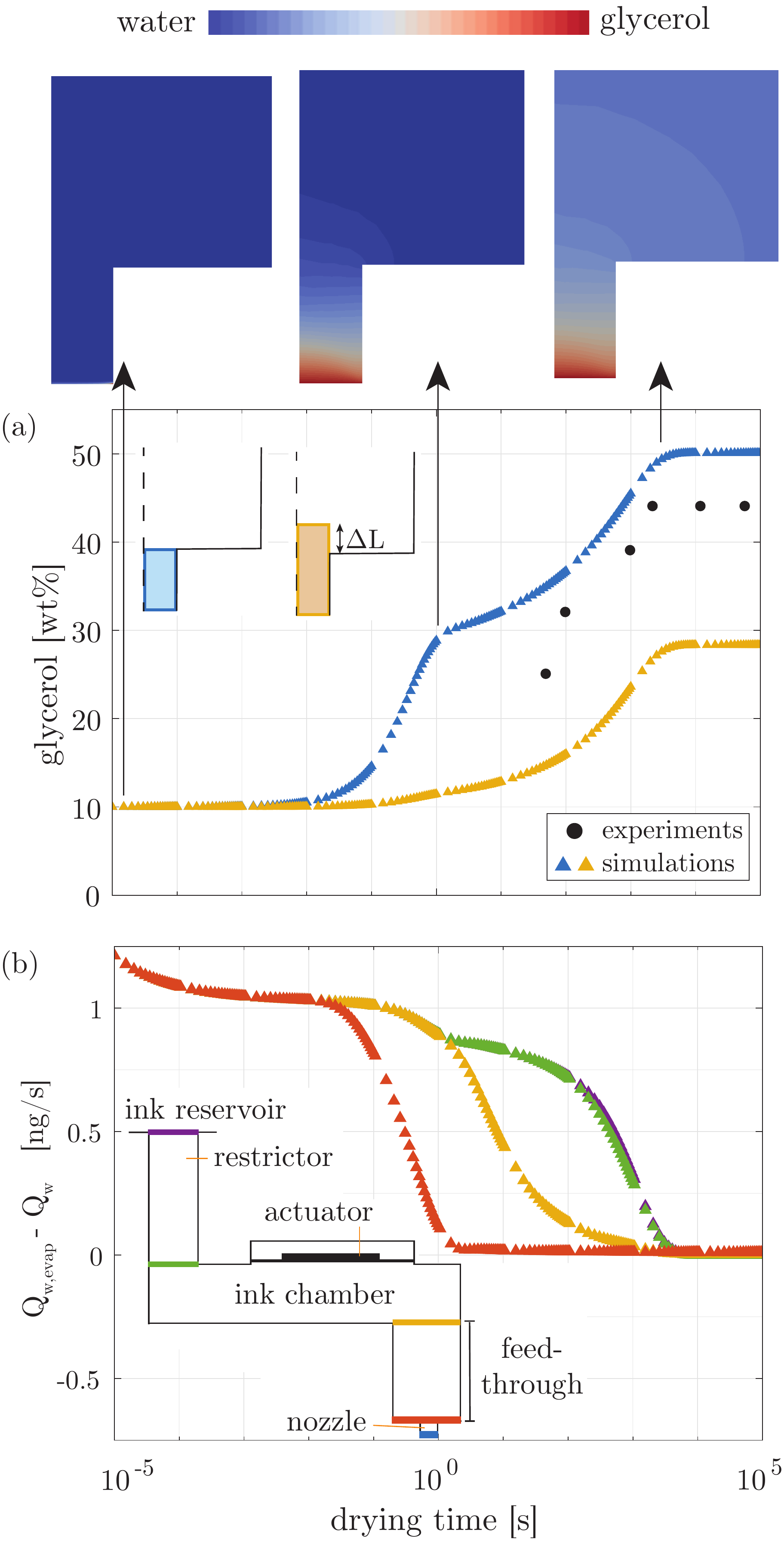}
	\caption{(a) Evolution of the glycerol concentration in the nozzle during drying time. The numerical results ($\triangle$) show the average glycerol concentration for two different nozzle lengths: the geometrical length of the nozzle (blue); and an effective nozzle length (yellow). The experimental ringdown frequency and decay rate are converted to a glycerol concentration in the nozzle using the analytical model (o). The numerical snapshots above the figure show the glycerol concentration in the nozzle and the feedthrough at the indicated drying times. (b) Results from the numerical simulations showing the difference between the flow rates of water out of the nozzle (through evaporation $Q_{w,evap}$) and through the boundaries between the other regions (through advection and diffusion $Q_w$). The different colors indicate the region boundaries as illustrated in the inset.}
	\label{fig:saturation}
\end{figure}

Next, the plateau that is observed in both the ringdown frequency and the decay rate after a drying time of \SI{2000}{\second} is investigated. The analytical model indicates that the average concentration in the nozzle is \SI{44}{\wtpercent} glycerol. The numerical model described in section~\ref{sec:num} is employed to further investigate this. To validate the numerical model, the ringdown frequency and decay rate that are obtained from increasing a homogeneous glycerol concentration in the nozzle in the simulations are also included in Fig.~\ref{fig:damp_vs_freq} (triangular markers). It can be seen that the numerical data agrees well with those from the analytical model. The simulations especially agree well with the experimental datapoints, which is the region of interest. The small deviation between the analytical and numerical models at high decay rates is expected to originate from the nozzle length that is not corrected for viscous friction in the analytical model, but only for its inertia~\cite{Dagan1982}.

The plateau is now investigated by simulating the evaporation process at 0~$\%$RH and extracting the glycerol concentration in the nozzle by taking the average value in the nozzle. Figure~\ref{fig:saturation}a presents two curves for the numerically obtained average glycerol concentration in the nozzle: one where the average concentration is taken from the geometrical area of the nozzle (blue) and one where the area is extended into the feedthrough by the length $\Delta L$ to account for the inertia of the flow (yellow datapoints, see section~\ref{sec:model}). Furthermore, a kink is visible in both numerical curves, at $\sim$~\SI{1}{\second}. As shown in the snapshots in Fig.~\ref{fig:saturation}a, the kink appears when the concentration gradient reaches the feedthrough. To explain: the diffusion in the nozzle can be approximated as one dimensional as long as the concentration gradient extends upward into the nozzle. When the concentration gradient reaches the feedthrough, the channel widens and therefore the diffusion becomes three dimensional (upward and sideways). This effectively creates a larger reservoir of water to diffuse into the nozzle, thereby resulting in a decreased rate at which the concentration of glycerol in the nozzle is increasing. After some time, the concentration gradient in the feedthrough reaches the channel walls, and the diffusion becomes steady and one-dimensional again. The time of the kink can be approximated by the mass transfer timescale for the water/glycerol mixture to travel the length of the nozzle in a solution of \SI{10}{\wtpercent} glycerol: $\tau_{n} \approx L_n^2/D = (\SI{15}{\micro\meter})^2/\SI{0.6e-9}{\meter\squared/\second} =  \SI{0.4}{\second}$. These results show that the experimentally observed plateau in glycerol concentration after long drying times ($>$\SI{2000}{\second}) is also obtained from the numerical simulations.

To also include the experimental results in Fig.~\ref{fig:saturation}a, the glycerol concentration in the nozzle is estimated from the experimentally measured ringdown frequency and decay rate using our analytical model. Both experiments and numerical simulations show an initial increase in glycerol concentration in the nozzle, eventually reaching a plateau after the same drying time. This indicates that the mass of water lost via evaporation at the nozzle exit is replenished through the transport of water upstream of the nozzle at the same rate, resulting in a steady state. This picture is confirmed by extracting the difference in flow rate of water between evaporation out of the nozzle $Q_{w,evap}$ and transport through the different regions in the printhead $Q_w$, see Fig.~\ref{fig:saturation}b. Initially, the water loss due to evaporation is faster than the water replenishment from the reservoir. As the local concentration of glycerol gradually increases, the water evaporation rate decreases, and so does the transport of water inside the printhead. At the moment the plateau is reached, the transport of water through each region is the same as the rate of evaporation. With the same approach for the kink at $\sim$~\SI{1}{\second}, the timescale can be estimated for when the concentration gradient has passed through the system: $\tau_{s} \approx L_{total}^2/D = (\SI{1.2}{\milli\meter})^2/\SI{0.6e-9}{\meter\squared/\second} =  \SI{2400}{\second}$. This suggests the observed plateau occurs as soon as the concentration gradient reaches the ink reservoir and again demonstrates that the observed plateau in glycerol concentration is the result of a steady state of the water flux through the system.

\subsection{Recovering the liquid composition}\label{subsec:recovery}

\begin{figure}
    \centering
    \includegraphics[width=\columnwidth]{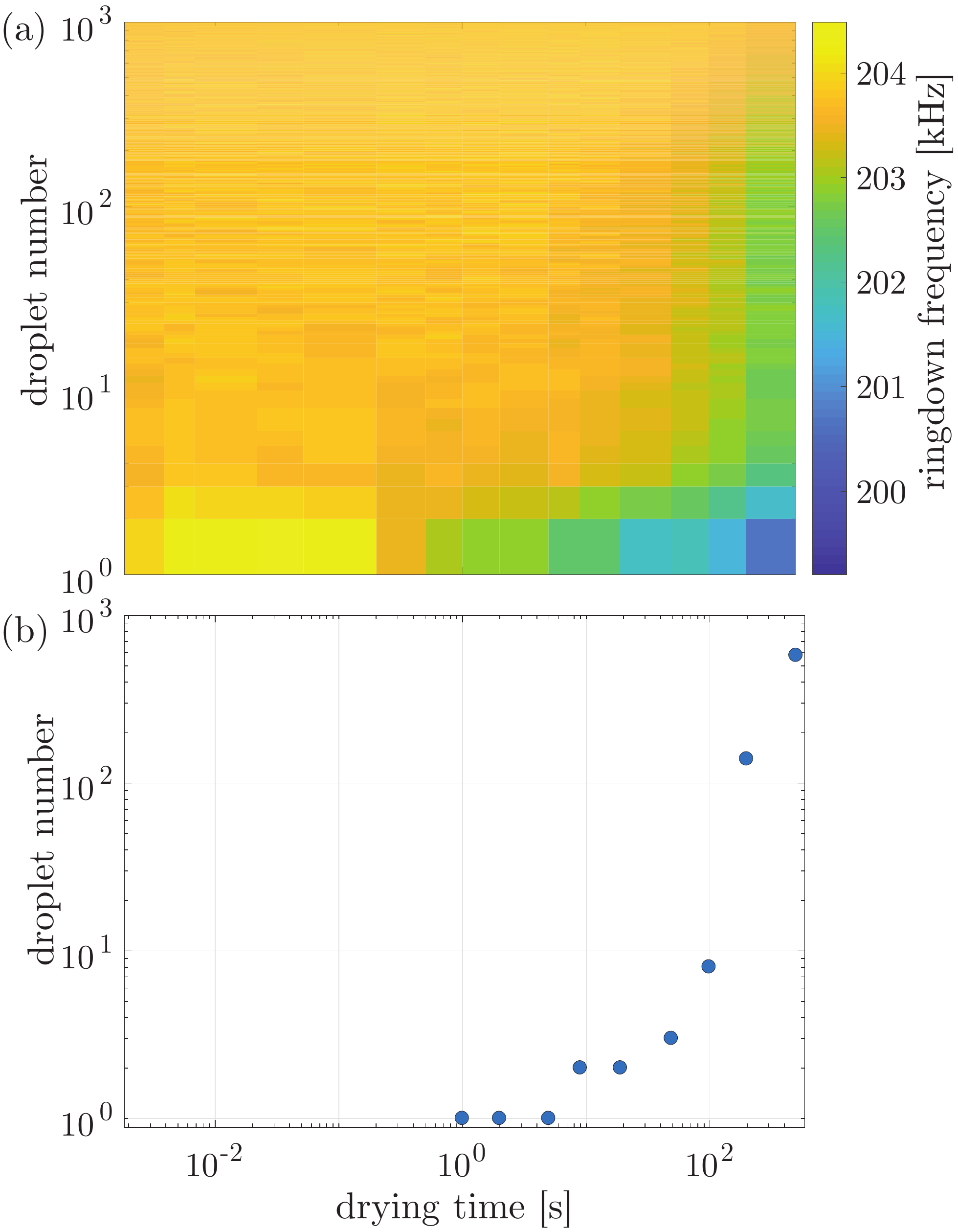}
    \caption{(a) The resonance frequency of the ringdown signal for the first 1000 droplets that form after different drying times. (b) The number of droplets that have to be jetted before the resonance frequency is back to within 0.5$\%$ of the resonance frequency without drying, on a log-log scale.}
    \label{fig:Result_return}
\end{figure}

Until now, the focus has been on the change in the liquid composition during the drying period. However, for all practical purposes, in the inkjet printer it is important to get back to the initial liquid composition to recover the well-controlled drop formation. Therefore, we now investigate the number of droplet formation events required to achieve the initial (pre-drying) liquid composition after a certain period of drying. To do so, the ringdown signals of the probe pulses of the first 1000 droplets after different drying times were measured. Figure~\ref{fig:Result_return}a displays the resonance frequency for the different drop numbers for different drying times. The lowest horizontal row shows the frequencies before the first droplets after the drying time, and is the same as the frequencies in Fig.~\ref{fig:Result_PAINt}a for 38~$\%$RH. At short drying times there is no variation in resonance frequency with drop number. With increasing drying time, more droplet formation events are required to return to the resonance frequency of the shorter drying times. This is evident from Fig.~\ref{fig:Result_return}b, which shows the number of droplet formations required for the resonance frequency to return to within 0.5$\%$ of the original resonance frequency. As can be observed from Fig.~\ref{fig:Result_return}b, the number of droplets required for the recovery increases more than exponentially with drying time. This can be the result of the concentration gradient in the liquid extending beyond the nozzle region for long drying times, where the flow is more complex during the jetting process, containing stagnant fluid and a vortex as shown in~\cite{fraters-2019-prappl}.

\section{Conclusion \& Outlook}

In drop-on-demand inkjet printing there can be loss in control over the print quality due to intermittent periods of no jetting during the printing process. During these idle periods, the nozzle can dry out and a multicomponent ink (in this case, a mixture of water and glycerol) can change in composition at the nozzle exit because of selective evaporation. By measuring the ringdown signal of the ink channel acoustics in response to a probe pulse, the change in ringdown frequency and decay rate are measured for different drying times. An analytical model is proposed, which enables the calculation of the glycerol concentration in the nozzle from the experimentally-measured ringdown frequency and decay rate. It was also observed that after an extended drying time, an equilibrium is reached between the amount of water lost at the nozzle exit through evaporation and the amount of water transported from the bulk to the nozzle exit via advection and diffusion. The result is a liquid composition that does not change with drying time anymore. Finally, the amount of jetted droplets required to recover from the drying process is observed to increase stronger than exponentially with drying time.

Being able to probe the local concentration of a multi-component mixture at the nozzle exit allows for the investigation of more complex problems, such as the influence of a variety of components typically used to control sessile droplet evaporation in inks, including particles and surfactant. Thus, the results presented here allow for a better understanding of the requirements to control the printing process, to non-invasively measure the change of concentration during evaporation, and to measure the effectiveness of mitigation strategies such as fluid mixing (non-jetting) pulses during drying periods, which we have observed to influence the concentration profile.

\vspace{-0.7 cm}
\section*{Acknowledgements}
This work was supported by an Industrial Partnership Programme of the Netherlands Organisation for Scientific Research (NWO), co-financed by Canon Production Printing Netherlands B.V., University of Twente, and Eindhoven University of Technology. Furthermore, T.S. also acknowledges financial support of the Max Planck Center Twente for Complex Fluid Dynamics.



%

\end{document}